\documentclass[twocolumn,showpacs,preprintnumbers,amsmath,amssymb,prb]{revtex4}
\usepackage{graphicx}
\usepackage{dcolumn}
\usepackage{bm}
\usepackage{amssymb}
\usepackage{amsmath}
\usepackage{epsf}
\usepackage{amssymb}
\usepackage{amsmath}
\usepackage{epsf}
\usepackage{graphics}
\usepackage{verbatim}
\begin{document}

\title{Electron spin decoherence in diluted magnetic quantum wells}

\author{P. M. Shmakov$^{2}$}
\author{A. P. Dmitriev$^{1,2}$}
\author{V. Yu. Kachorovskii$^{1,2}$}
\affiliation{
$^{1}$Institut f\"ur Nanotechnologie, Forschungszentrum Karlsruhe,
76021 Karlsruhe, Germany
\\
$^{2}$A.F.Ioffe Physico-Technical Institute,
194021 St.~Petersburg, Russia
}

\date{\today}

\begin{abstract}
{We study electron spin dynamics in   diluted magnetic quantum wells. The electrons are coupled by
exchange interaction
 with randomly distributed magnetic ions polarized by magnetic field $\mathbf{B}.$  This coupling  leads to both spin relaxation and spin decoherence.
We demonstrate that even very small spatial fluctuations of quantum well width dramatically increase rate of decoherence.
Depending on the strength of exchange interaction and the amplitude of the fluctuations the decoherence can be homogeneous or inhomogeneous.
In the homogeneous regime,     the transverse
(with respect to $\mathbf{B}$) component of electron spin  decays on the short time scale exponentialy  
while the long-time spin dynamics is  non-exponential demonstrating long-lived power law tail.  In the inhomogeneous case, the transverse spin component decays exponentially with the exponent quadratic in time.
}
\end{abstract}

\maketitle

Effective manipulation of  spin degree of freedom  in a semiconductor device by external electric and magnetic fields is one of the primary goals of   spintronics. \cite{avsh} A possible way to increase coupling of electron spin  to  external magnetic  field is to use semiconductor-based materials, which incorporate magnetic elements. Such materials  are called  diluted magnetic semiconductors (DMS). The most common DMS are II-VI and IV-VI compounds with magnetic impurities (usually, Mn),  and also  III-V crystals, \cite{4,5} like  Ga$_{ {1-x}}$Mn$_{ x}$As  (typically, $x< 10\%$). The DMS  combine magnetic and semiconductor properties in a single material. This offers a large prospect for  applications. In particular, DMS are considered to be the most promising candidates  in creating room-temperature ferromagnetic systems, which can be easily manipulated as semiconductors.  Semiconductor heterostructures doped by magnetic impurities have already demonstrated  exciting physical phenomena specific for magnetic systems: coherent spin excitations, \cite{a3,a4} magnetic polaron formation, \cite{a5}  ferromagnetic hole alignment, \cite{a6} etc.

 Many remarkable features of DMS, such as the large Zeeman splitting of the electronic bands and the giant Faraday rotation, are induced by the exchange interaction between the localized electrons on d-shells of Mn ions and delocalized band carriers (see Ref.~\onlinecite{dyak} for review). This interaction  is also  responsible for the collective nature of spin excitations in DMS, arising of novel collective modes, \cite{6,8}, anticrossing of the electron and ion spin precession
 frequencies, \cite{6,1} spin susceptibility enhancement, \cite{7} etc.

The fluctuations of exchange field around the average value lead to  electron spin relaxation  and spin decoherence with characteristic times $T_1$  and $T_2,$ respectively. Recently, the decoherence rate was measured in Cd$_{1-x}$Mn$_x$Te two-dimensional (2D) structure. \cite{2}  The observed $T_2$ was found to be at least
one order magnitude shorter than the  decoherence time   predicted theoretically in Ref.~\onlinecite{3},
where  fluctuations of exchange field were linked to delta-correlated fluctuations of magnetic ion concentration.
In this paper, we demonstrate that even very small fluctuations of quantum well width dramatically increase rate of decoherence. Our estimates show that this mechanism can explain short decoherence time observed in Ref.~\onlinecite{2}.

We consider the 2D degenerate electron gas located in the $(x,y)$ plane interacting with the magnetic ions randomly distributed with average  2D concentration $n_J$,
and doping profile $f(z)$ in the growth direction $(\int f(z)dz = 1)$.  The system is placed
into  the magnetic field which we assume to be  parallel to the well plane ($\mathbf B \parallel \mathbf e_x$) as it was the case in Ref.~\onlinecite{2}. The field
 leads to Zeeman splitting of both electron and ion spin levels with energies $\hbar \omega_e$ and $\hbar \omega_J,$ respectively.
 The Hamiltonian of the system is given by
\begin{align}
&\hat H = \hat H_e +
\hat H_J +
\hat H_{Je},
\\
&\hat H_e = \frac{\mathbf p^2}{2m} + U(\mathbf{r}) + \hbar \omega_e \hat s_x ,
\\
&\hat H_J = \hbar \omega_J \sum_k{\hat J_{kx}},
\\
&\hat H_{Je} = \alpha \mathbf{\hat s}\sum_k{\mathbf{\hat J}_k \delta(\mathbf r - \mathbf R_k)}|\Psi(z_k)|^2.
\end{align}
Here $\hat H_e$ is the Hamiltonian of an electron in  the external magnetic field, $\mathbf r=(x,y)$ and $\mathbf p=(p_x,p_y)$ are respectively
 electron in-plane position vector and momentum,  $U(\mathbf r)$ is the random impurity potential, which we assume to be short-range,
$\hat H_J$ is the Hamiltonian of the ions and $\hat H_{Je}$ represents the exchange interaction
between spin $\mathbf s$ of an electron   placed at the lowest level in the well (with the wave function $\Psi(z)$)  and the spins $\mathbf J_k$ of the ions located at points $(\mathbf R_k,z_k)$.
The strength of the interaction is characterized by   constant $\alpha.$

It is convinient to rewrite $\hat H_{Je}$ as
\begin{equation}
\hat H_{Je} = \langle\hat H_{Je}\rangle + \delta\hat H_{Je},
\end{equation}
where the angular brackets mean averaging over ion positions and thermal averaging and  $\delta\hat H_{Je}$ describes fluctuations of exchange field.
The term $\langle\hat H_{Je}\rangle$ leads to significant renormalization of the electron spin precession frequency  \begin{equation}
\label{omega}
\omega = \omega_e +\frac{\alpha \langle J_x \rangle n_J}{\hbar} \int f(z)|\Psi(z)|^2dz,
\end{equation}
Typically,  $\omega \gg \omega_e$,\cite{1,2} which means that the electron spin precession frequency  is mostly determined by the effective magnetic field, created by polarized ions.

The fluctuations of the exchage  field arising due to random distribution of $(R_k,z_k)$
lead both to electron spin relaxation and to decoherence.  The analysis of these processes
 shows \cite{3}
 that in 2D case the  longitudinal and transverse (with respect to $\mathbf B$)  components of the electron spin decay exponentially with the characteristic times $T_1^\prime \sim T_2^\prime \sim  {2a^2\hbar^3}/ {3\alpha^2 m n_J }.$
Comparing the results  of Ref.~\onlinecite{3} with the recent experimental data, \cite{2} one can see that experimentally observed decoherence time $T_2$ is much shorter (by about an order of magnitude) than $T_2'$, which implies that  delta-correlated density fluctuations can not provide sufficient fluctuations of the exchange field.

Below we demonstrate  that spatial fluctuations of quantum well width can (at certain conditions)  lead to significantly shorter $T_2$. Physically, this happens because  such
 fluctuations induce the long-range fluctuations of the effective magnetic field acting on the electron spin.

 First, we notice
  that the  effective magnetic field induced by exchange interaction depends on the doping profile $f(z)$ and quantum
   well geometry (see  Eq.~\eqref{omega}).  Expressions for $T_1'$ and $T_2'$  were derived \cite{3} for the case
   of homogeneous distribution of magnetic ion and infinitely deep quantum well, having constant width $a_0.$
Let us  now assume, for example, that the quantum well is infinitely deep and magnetic ions concentrate close to the center of the well
(these assumptions more or less correspond to the experimental situation \cite{2}).
Assuming also that well width slightly fluctuates, $a(\mathbf r)=a_0+\delta a(\mathbf r)$
we find that spin precession frequency becomes $\mathbf r-$dependent:
 \begin{equation}
\omega \rightarrow \omega (\mathbf r)=\omega_e+ \frac{2\alpha \langle J_x \rangle  n_J}{\hbar a( \mathbf r)}=\omega_0+\delta \omega(\mathbf r),
\end{equation}
where $\omega_0=\omega_e+{2\alpha \langle J_x \rangle  n_J}/{\hbar a_0},$ $\delta \omega(\mathbf r) \approx -{2\alpha \langle J_x \rangle  n_J \delta a(\mathbf r)}/{\hbar a_0^2}.$
We assume the fluctuations
to be Gaussian with a spatial scale $d$:
\begin{equation}
\langle\delta \omega(\mathbf{r})\delta \omega(\mathbf{r'})\rangle =
2\omega^2_*\,\chi\left(\frac{|\mathbf r - \mathbf r^\prime|}{d}\right).
\label{delta-delta}
\end{equation}
Here
 $\omega_*^2={2\alpha^2 \langle J_x \rangle^2  n_J^2}{\langle \delta a^2(\mathbf r) \rangle}/\hbar^2 {a_0^4}$ is  the amplitude of the flucuations
  and $\chi(\xi)$ is a dimensionless function
 ($\chi(0)=1,~ \chi(\xi) \to 0,~{\rm for}~ \xi \gg 1 $).  As seen from Eq.~\eqref{delta-delta}, $\langle\delta \omega(\mathbf{r})\delta \omega(\mathbf{r'})\rangle \sim \langle J_x \rangle^2,  $ so that at low magnetic fields the fluctuations vanish and  decoherence is due to the mechanism proposed in Ref.~\onlinecite{3}. However, at relatively strong fields corresponding to almost full polarization of ions (this was the case in the experiment \cite{2})   the fluctuations are sufficiently large and  dominate the  decoherence.

  We will consider time evolution of the spin excitations concentrated near Fermi surface.
  The total transverse spin (per unit area) can be written as
  $S_+(t)=S_y + iS_z = e^{i\omega_0 t} S(t),$ where $S(t)=\int s_+(\mathbf r,\varphi,t)d \varphi d^2\mathbf r/\Omega  $ is slowly decaying amplitude,  $\Omega$ is the sample area
  and $s_+= s_y + is_z$ is the spin density that obeys 
    the following quasiclassical kinetic equation \cite{dyakonov_old,dyakonov_new} 
 \begin{equation}
\frac{\partial s_+}{\partial t} + v_F\mathbf{n}\frac{\partial s_+}{\partial \mathbf{r}} - i\delta \omega(\mathbf{r}) s_+ = \mathrm{{St}}\{ s_+\},  \label{kin} \end{equation}
where $\varphi$ is the velocity angle, $v_F$ is the Fermi velocity $\mathbf n=\mathbf e_x \cos \varphi +\mathbf e_y \sin \varphi,$
  and $\mathrm{St}\{s_+\}$ is the collision integral describing the elastic  scattering on the impurity potential $U(\mathbf r)$ with the mean free path $l$.   The quasiclassical approach based on Eq.~\eqref{kin}  is valid  provided that  $d \gg \hbar/mv_F $ and $\hbar \omega(\mathbf r) \ll mv_F^2.$ Below we will solve this equation with the  initial condition $s_+(\mathbf r,\varphi,0)=1.$  

    The spin decoherence can be homogeneous or inhomogeneous  depending on the parameter $\omega_* \tau, $
    where $\tau$ is a characteristic time  required  for electron to travel the distance of the order of $d.$
    For $d\ll l$, this time is given by $\tau =  d/v_F,$ while for $d \gg l$, $\tau = d^2/4D \sim d^2/l v_F,$  where $D =  v_F l/2$ is the diffusion coefficient.

   For $\omega_* \tau \gg 1 ,$ electron spins in different correlation regions rotate independently with local frequencies. Hence, the  decoherence is
    inhomogeneous
          and the  transverse spin decays as
           \begin{equation}
 S(t) \approx  \langle e^{i\delta \omega(\mathbf r)t } \rangle= e^{-{\omega_*^2 t^2}}.
 \end{equation}

       In the opposite case, $\omega_* \tau \ll 1 ,$ electron is visiting  many correlated regions during decoherence time and the   decoherence is
    homogeneous.    First, we assume $d \ll l.$  If the inequality $T_2 v_F \ll l$ is also satisfied,
          the electron motion on the time scale on the order of $T_2$ is ballistic. Neglecting the collision integral in Eq.~\eqref{kin}, we obtain
        \begin{align}
 &S(t) = \left\langle e^{i \int_0^t \delta \omega (v_F \mathbf n \tau)\mathrm d \tau}\right\rangle 
 \label{ball}
 \\
 \nonumber
  &= e^{-\omega_*^2 \int_0^t \int_0^t \chi (v_F |\tau_1-\tau_2|/d)
 \mathrm d \tau_1\mathrm d \tau_2}.
  \end{align}
  For $t \gg d/v_F$, Eq.~\eqref{ball} becomes
 \begin{equation}
 S(t) = e^{- t/T_2}, ~ \frac{1}{T_2}=\frac{2\omega_*^2 d}{v_F}\int\limits_0^\infty{\chi(\xi)\mathrm d \xi}.
 \label{T2-1}
 \end{equation}
    It turns out  that Eq.~\eqref{T2-1} is also valid for $T_2 v_F \gg l.$
   To see this, one can iterate Eq.~\eqref{kin} with respect to small $\delta \omega$ and decouple correlations, which yields  $ 1/T_2=\int d\mathbf r' \int dt'\langle \delta\omega(\mathbf r)G(\mathbf r-\mathbf r', t-t') \delta \omega(\mathbf r')\rangle.    $  Here $G(\mathbf r,t)$ is the Green function of  Eq.~\eqref{kin} with $\delta \omega(\mathbf r)=0 $ (Green function of the Boltzmann equation)
   integrated  over initial velocity directions and averaged over final velocity directions. This function
   can be presented as a sum over processes with different
   number of collisions:   $G(\mathbf r,t)=\langle \delta (\mathbf r-v_F \mathbf n t)\rangle_{\varphi} \exp(-tv_F/l) +\sum_{N=1}^\infty G_N(\mathbf r,t),$
   where the first term is the ballistic contribution,
   which gives  Eq.~\eqref{T2-1}. Relative contribution of other terms to $1/T_2$
   (compared to Eq.~\eqref{T2-1}) may be shown to be on the order of parameter $(d/l)\ln(1/\omega_*\tau) \ll 1$.\cite{com}

       For              $d \gg l$ and $ \omega_*\tau =\omega_* d^2/4D\ll 1,$
             the decoherence is also homogeneous. In this case the electron motion is diffusive and  one can by standard means reduce Eq.~\eqref{kin} to diffusion equation:
\begin{equation}
\frac{\partial s_+}{\partial t} - D\Delta s_+ - i\delta \omega(\mathbf{r}) s_+ = 0.
\label{diff2}
\end{equation}
Iterating Eq.~\eqref{diff2} with respect to  $\delta \omega$ and decoupling correlations, we obtain
\begin{equation}
\frac{\partial s_+}{\partial t} - D\Delta s_+
+ \int d\mathbf r' dt' \Lambda_{\mathbf r - \mathbf r', t-t'} s_+(\mathbf r',t')=0,
\label{diff3}
\end{equation}
where
\begin{equation}
\Lambda_{\mathbf r - \mathbf r', t-t'}=
2\omega^2_* \chi\left(\frac{|\mathbf r - \mathbf r^\prime|}{d}\right)
G_D(\mathbf r - \mathbf r', t-t'),
\nonumber
\end{equation}
and $\hat{ G}_D= (\partial/\partial t - D\Delta)^{-1}$ is the Green function of the diffusion equation. Solving Eq.~\eqref{diff3} with the initial condition $s_+(\mathbf r,\varphi,0)=1$, we find
 \begin{equation}
S(t) = \int_{-\infty}^{\infty}{ \frac{d\omega}{2\pi} \frac{e^{-i\omega t}}{- i\omega + 2\omega_*^2 d^2 \int_0^{\infty} \frac{\tilde\chi(qd)}{Dq^2-i\omega} {\frac{q dq}{2\pi}  }}},
\label{intdiff}
\end{equation}
where $\tilde\chi$ is the Fourier transform of $\chi$.
\begin{figure}[htt!]
 \leavevmode \epsfxsize=5.5cm \centering{\epsfbox{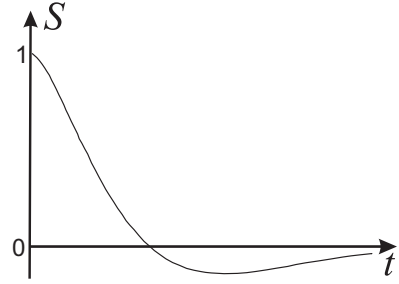}}
\caption{ Decay of the transverse spin amplitude in the diffusive case }
\end{figure}
The integrand in Eq.~\eqref{intdiff}  has two  poles and a branch cut along the negative imaginary axis. The poles are at the points
$\omega_{1,2} \approx  ({2 \tilde\chi(0)\omega_*^2 \tau}/{\pi})\left[\pm\pi-2i \ln \left({1}/{\omega_*\tau}\right)\right],$
where $\tau= d^2/4D, \tilde\chi(0)=2\pi\int_0^\infty \xi \chi(\xi) d\xi$.

\begin{figure}[ht!]
 \leavevmode \epsfxsize=5.5cm \centering{\epsfbox{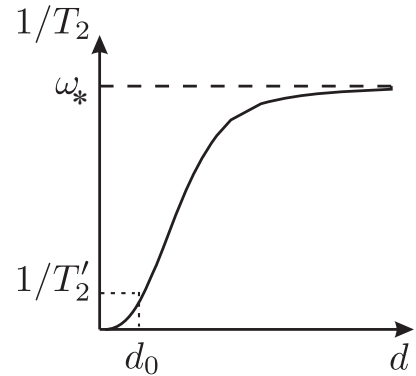}}
\caption{Decoherence rate as a function of the correlation radius of the fluctuations.
For $d>d_0$ mechanism related to fluctuations of the well width dominates   }
\end{figure}

The imaginary parts of the poles
are large compared to the real ones, so that the poles contribution  is approximately given by

\begin{equation} S(t) \approx 
\exp\left[-\frac{4\tilde\chi(0)\omega_*^2  \tau t}{ \pi}\ln\left(\frac{1 }{\omega_* \tau}\right)\right].
\label{overdamped} \end{equation}
This contribution dominates at short times,
$\omega_*^2 \tau t \ll 1 .$
For  $\omega_*^2 \tau t \gg 1, $ the main contribution is due to branch cut, yielding
\begin{equation} S(t) \approx
-\frac{\pi}{2\tilde\chi(0)}\frac{1}{\omega_*^2 \tau t \,\ln^2\left( t/\tau\right)}.
\label{asympt} \end{equation}

The contribution of the branch cut is negative.
 From Eqs.~\eqref{overdamped} and \eqref{asympt}
we see that the amplitude $S(t)$  changes sign as shown in Fig.~1. We also see that
a long-lived power-law tail appears in the transverse spin polarization.

 Above we assumed that $\mathbf B$ is parallel to the well plane and neglected the effect of the field on the orbital motion. The results are also valid for $\mathbf B \parallel \mathbf e_z$ provided that
  $R_c \gg l$ ($R_c$ is the cyclotron radius).  Our calculations can be easily  generalized for  the opposite case, $R_c \ll l$. In particular,  in the homogeneous ballistic regime, under the assumptions $\omega_*\tau \ll 1, d \ll R_c \ll l, R_c \ll T_2v_F\ll l$, the transverse spin can be calculated in analogy with Eq.~\eqref{ball} by averaging of decoherence action calculated along ballistic trajectory. The result looks \cite{com1}
\begin{equation}
 S(t) = e^{- t^2/T_2^2}, ~ \frac{1}{T_2^2}=\frac{\omega_*^2 d}{\pi R_c}\int\limits_0^\infty{\chi(\xi)\mathrm d\xi}.
 \end{equation}

 From  equations derived above
 one can  see  that increasing correlation radius $d$ decreases $T_2$ both for $d \ll l$ and for $d \gg l$. The maximal value of $1/T_2$ is on the order of $\omega_*.$
One can show that $\omega_* T_2^\prime \gg 1$  for typical values of parameters,\cite{2} which implies that suggested mechanism  dominates already at small $d$  ($d > d_0$, see Fig. 2)
and might be responsible for  short values of decoherence time observed in the experiment.\cite{2}
    The parameter $d_0$  is estimated as follows:
    $
    d_0 \sim {k_F a^2}/{n_J \delta a_0^2},
    $
    for $l \gg d$  and
    $
    d_0 \sim \sqrt{{k_F a^2 l}/{n_J \delta a_0^2},}
    $
    for  $l \ll d$ regime. Taking $\delta a_0=\sqrt{\langle \delta a^2(\mathbf r) \rangle}$ on the order of the lattice constant,  for typical values of experimental parameters~\cite{2} we find that $d_0 \approx 10 nm$.
\newline
\indent We thank M.~Glazov for useful discussion. The work was  supported by RFBR, by
grant of Russian Scientific School, by the Grant of Rosnauka (project number 02.740.11.5072) and by programmes of the RAS.
 V.Yu.K. was  supported by  Dynasty foundation.


\begin{thebibliography}{8}
\bibitem{com} We do not consider here intermediate case $d/l <1,$ $(d/l)\ln (1/\omega_*\tau) >1.$
\bibitem{com1} The  quadratic dependence of decoherence exponent on time (decoherence acceleration) is  analogous
to    the acceleration of the spin relaxation in a system with random spin-orbit coupling.  \cite{glazov}
\end{thebibliography}
\end{document}